\documentclass[10pt,twocolumn,preprintnumbers,prd,superscriptaddress,nofootinbib,floatfix,showpacs,showkeys,aps]{revtex4-2}

\usepackage{float}
\usepackage{graphicx}  
\usepackage{multirow}
\usepackage{natbib}
\usepackage{pifont}
\usepackage{amssymb}
\usepackage{bm}        
\usepackage{amsfonts}  
\usepackage{amsmath}   
\usepackage{amssymb}   
\usepackage[titletoc]{appendix}
\usepackage{color}
\usepackage{hyperref}
\usepackage{cleveref}
\usepackage[english]{babel}
\usepackage{booktabs}
\usepackage{mathtools}
\usepackage{subfigure}
\usepackage{enumerate}
\usepackage{soul}
\usepackage{xurl}
\usepackage{hyperref}

\definecolor{vividviolet}{rgb}{0.62, 0.0, 1.0}
\definecolor{amaranth}{rgb}{0.9, 0.17, 0.31}
\definecolor{palatinateblue}{rgb}{0.15, 0.23, 0.89}
\definecolor{brightpink}{rgb}{1.0, 0.0, 0.5}
\definecolor{cornflowerblue}{rgb}{0.39, 0.58, 0.93}
\definecolor{deepcarminepink}{rgb}{0.94, 0.19, 0.22}
\definecolor{radicalred}{rgb}{1.0, 0.21, 0.37}

\hypersetup{ linktoc=all,
    colorlinks, linkcolor={palatinateblue},
    citecolor={brightpink}, urlcolor={amaranth}
}

\begin{document}

\title{Gravitational waves from two scalar fields unifying the dark sector with inflation}

\author{Orlando Luongo}
\email{orlando.luongo@unicam.it}
\affiliation{University of Camerino, Via Madonna delle Carceri, Camerino, 62032, Italy}
\affiliation{Istituto Nazionale di Fisica Nucleare (INFN), Sezione di Perugia, Perugia, 06123, Italy}
\affiliation{Department of Mathematics and Physics, SUNY Polytechnic Institute, Utica, NY 13502, USA}
\affiliation{INAF - Osservatorio Astronomico di Brera, Milano, Italy}
\affiliation{Al-Farabi Kazakh National University, Almaty, 050040, Kazakhstan}

\author{Tommaso Mengoni}
\email{tommaso.mengoni@unicam.it}
\affiliation{University of Camerino, Via Madonna delle Carceri, Camerino, 62032, Italy}
\affiliation{Istituto Nazionale di Fisica Nucleare (INFN), Sezione di Perugia, Perugia, 06123, Italy}
\affiliation{INAF - Osservatorio Astronomico di Brera, Milano, Italy}

\author{Paulo M. S\'a}
\email{pmsa@ualg.pt}
\affiliation{Departamento de F\'\i sica, Faculdade de Ci\^encias e Tecnologia, Universidade do Algarve, Campus de Gambelas, 8005-139 Faro, Portugal}
\affiliation{Centro de Investiga\c{c}\~ao e Desenvolvimento em Matem\'atica e Aplica\c{c}\~oes (CIDMA), Universidade de Aveiro, Campus de Santiago, 3810-193 Aveiro, Portugal}

\begin{abstract}
We investigate the gravitational-wave background predicted by a two-scalar-field cosmological model that aims to unify primordial inflation with the dark sector, namely late-time dark energy and dark matter, in a single and self-consistent theoretical framework. The model is constructed from an action inspired by several extensions of general relativity and string-inspired scenarios and features a non-minimal interaction between the two scalar fields, while both remain minimally coupled to gravity. In this context, we derive the gravitational-wave energy spectrum over wavelengths ranging from today’s Hubble horizon to those at the end of inflation. We employ the continuous Bogoliubov coefficient formalism, originally introduced to describe particle creation in an expanding Universe, in analogy to the well-established mechanism of gravitational particle production and, in particular, generalized to gravitons. Using this method, which enables an accurate description of graviton creation across all cosmological epochs, we find that inflation provides the dominant gravitational-wave contribution, while subdominant features arise at the inflation-radiation, radiation-matter, and matter-dark energy transitions, i.e., epochs naturally encoded inside our scalar field picture. The resulting energy density spectrum is thus compared with the sensitivity curves of the planned next-generation ground- and space-based gravitational-wave observatories. The comparison identifies frequency bands where the predicted signal could be probed, providing those windows associated with potentially detectable signals, bounded by our analyses. Consequences of our recipe are thus compared with numerical outcomes and the corresponding physical properties discussed in detail.
\end{abstract}




\maketitle 
\tableofcontents


\section{Introduction}

Gravitational waves, predicted by Einstein in 1916 \cite{Einstein:1916}, were only directly detected almost a hundred years later, in 2015, when the two Laser Interferometer Gravitational-Wave Observatory (LIGO) detectors simultaneously observed a transient gravitational wave signal corresponding to the merger of two massive black holes \cite{Abbott:2016}.

The importance of this breathtaking advance cannot be underestimated, as it not only represents a new test of the general theory of relativity regarding the existence of gravitational waves and binary stellar-mass black hole systems, but also opens a new window on the Universe, \emph{inaugurating the era of gravitational-wave astronomy} \cite{Bailes:2021}.

Among the multitude of possible sources of gravitational waves, those of primordial origin appear particularly important. 
These are produced across early stages of the Universe's evolution and, thus, may carry information about the inflationary period and the transition to the radiation-dominated epoch (for a recent review, see e.g.~Ref.~\cite{Roshan:2025}).

Considerable effort is being devoted to the creation of the next generation of ground- and space-based gravitational wave detectors, which, operating over a wide range of frequencies with enhanced sensitivity, can reveal this information, shedding light on the physics of the early Universe \cite{Lisa}.

Since the striking discovery that the Universe is presently undergoing a second period of accelerated expansion \cite{Riess:1998, Perlmutter:1999}, driven by a still unknown form of energy, dubbed dark energy, several attempts have been made to unify within a single theoretical framework both inflation and this energy.
If dark matter---whose nature remains currently elusive---is also included, then one can imagine that both the dark sector, composed of dark energy and dark matter, plus inflation, may be framed in terms of a \emph{triple unification cosmological scheme} \cite{Odintsov, o1, Liddle:2008, Henriques:2009, o2, o3, Paliathanasis:2023moe, o4}.

Among all possibilities, a relevant unification scheme was proposed by virtue of a two-scalar-field cosmological model \cite{sa-2020a, Sa:2020}, whose action was inspired by a variety of gravity theories, such as the generalized hybrid metric-Palatini, the Jordan-Brans-Dicke, Kaluza-Klein, $f(R)$, and string theories. 
The so-obtained unification scheme depends on the non-minimal coupling between the two scalar fields, which, however, are minimally coupled to gravity.
The model is defined by virtue of an unspecified \emph{a priori} potential, with one free kinetic energy term and another coupled with the other scalar field. 

Motivated by the above considerations, in this work, we calculate the full gravitational wave energy spectrum arising within this cosmological model.
To do so, we compute the spectrum from the minimum to the maximum allowed frequencies, corresponding to wavelengths equal to the Hubble distance today and at the end of the inflationary period,  respectively.
We start by postulating the effective potential of one of the two scalar fields, then proceed to compute the equations of motion and the corresponding Bogoliubov coefficients, used as a formalism toward the gravitational wave determination.
In particular, adopting the continuous Bogoliubov coefficients method is motivated and preferred since, once the dynamics of the Universe is known, it provides a straightforward way to compute the full spectrum of gravitational waves within a single framework, naturally avoiding issues related to the overproduction of gravitons at high frequencies and without imposing any cut-off ansatz. 
Accordingly, our findings are compared with the  sensitivity curves of the planned next-generation ground- and space-based gravitational-wave
detectors, showing those frequency windows in which our predictions are expected to be significant.
 
This article is organized as follows.
In Sec.~\ref{sezione2}, we briefly describe the two-scalar-field cosmological model unifying inflation, dark matter, and dark energy, while in Sec.~\ref{sezione3} we outline the formalism of continuous Bogoliubov coefficients used to calculate the corresponding gravitational-wave spectral energy density parameter as a function of frequency.
In Sec.~\ref{GW spectrum}, we present the spectrum and compare it with the sensitivity curves of the planned next-generation gravitational-wave detectors.
We also compare the spectra obtained for different values of the model's parameters.
Finally, in Sec.~\ref{conclusions}, we present our conclusions and perspectives.


\section{Two-scalar-field cosmological model} \label{2SF model}\label{sezione2}

As shown in Ref.~\cite{Sa:2020}, a triple unification of inflation, dark energy, and dark matter can be achieved within a two-scalar-field model given by the action 
\begin{align}
 S = {} & \int d^4x \sqrt{-g} \bigg[
  \frac{R}{2\kappa^2} - \frac12 (\nabla \phi)^2
    \nonumber
  \\
  & - \frac12 e^{-\alpha\kappa\phi} (\nabla \xi)^2
  - e^{-\beta\kappa\phi} V(\xi) \bigg].
 \label{action-2SF}
\end{align}
In the above expression, $g$ is the determinant of the metric $g_{\mu\nu}$, $R$ is the Ricci scalar, $\phi$ and $\xi$ are scalar fields, and $\alpha$ and $\beta$ are dimensionless parameters\footnote{Here, and in what follows, we use the notation $\kappa\equiv \sqrt{8\pi G}= \sqrt{8\pi}/m_\texttt{P}$, where $G$ is the gravitational constant and $m_\texttt{P} = 1.22 \times 10^{19}\, {\rm GeV}$ is the Planck mass.}. 

The potential $V(\xi)$ is, in principle, unspecified; the simplest choice providing such a triple unification has the form~\cite{Sa:2020}
\begin{equation}
V(\xi)=V_a+\frac12 m^2 \xi^2,
\end{equation}
where $V_a$ and $m$ are constants.
Choosing the above form of $V(\xi)$ is motivated by the simplest quadratic potential that does not alter the underlying physics, for example, inducing a phase transition, but rather introduces a bare mass term, associated with the scalar field.
This choice resembles the chaotic inflationary paradigm, where a simple massive scalar field is introduced. 

A detailed analysis of this cosmological model has been performed and can be found in Refs.~\cite{Sa:2020, Sa:2021}; here, we outline its main features, as follows below. 

Inflation is assumed to be of the warm type \cite{Berera:1995, kamali:2023, Berera:2023}, with the scalar field $\xi$ playing the role of the inflaton.
A preexisting radiation bath, with energy density $\rho_{\texttt{R}}$, is not diluted during the inflationary period due to a constant influx of energy from the inflaton (and also from the scalar field $\phi$).
The dissipation coefficients, which mediate this energy transfer, are chosen to depend on the temperature.

At a certain point, the energy density of radiation becomes greater than the energy density of the inflaton field, marking the end of the inflationary period. 
During the smooth transition to a radiation-dominated era, the dissipation coefficients are exponentially suppressed and quickly become negligible.

At the beginning of the radiation-dominated era, the inflaton $\xi$, now free from interaction with radiation, begins to oscillate rapidly around the minimum of its quadratic potential, thus behaving on average as a non-relativistic pressureless fluid, i.e., as cold dark matter.

After the radiation-dominated era, which includes the primordial nucleosynthesis period, cold dark matter, along with ordinary baryonic matter, begins to dominate the Universe's dynamics.
This leads to a matter-dominated era, which is long enough to allow for the formation of the observed large-scale structure of the Universe.

Finally, the scalar field $\phi$, which plays the role of dark energy, emerges as the dominant component of the Universe, giving rise to the current era of accelerated expansion.

In order to describe the Universe dynamics within a single framework, we assume a flat Friedmann-Lema\^{i}tre-Robertson-Walker metric 
\begin{equation}\label{metric}
 ds^2 = -dt^2 + a^2(t) d\Sigma^2, 
\end{equation}
where $a(t)$ is the scale factor, $t$ is the cosmic time, and $d\Sigma^2$ is the metric of the three-dimensional Euclidean space.
In this way, the equations of motion during the first stage of evolution (inflation and smooth transition to a radiation-dominated era) are
\begin{subequations} \label{SODE-1s}
\begin{align}
  &  \hspace{-0.5mm}  \xi_{uu} = - \bigg\{
     \bigg[ \frac{\ddot{a}}{a} + 2 \bigg( \frac{\dot{a}}{a} \bigg)^2
           +\frac{\dot{a}}{a} \Gamma_\xi e^{\alpha\kappa \phi} \bigg]\xi_u
    \nonumber
  \\
  & \hspace{8.5mm}
    - \alpha\kappa  \bigg( \frac{\dot{a}}{a} \bigg)^2 \phi_u \xi_u
    + m^2 \xi e^{(\alpha-\beta)\kappa \phi}
  \bigg\} \bigg( \frac{\dot{a}}{a} \bigg)^{-2},
      \label{Eq-xi-s1}
  \\
  & \hspace{-1.0mm}\phi_{uu} =  - \bigg\{
     \bigg[ \frac{\ddot{a}}{a} + 2 \bigg( \frac{\dot{a}}{a} \bigg)^2
           +\frac{\dot{a}}{a} \Gamma_\phi
     \bigg]\phi_u
  + \frac{\alpha\kappa}{2} \bigg( \frac{\dot{a}}{a} \bigg)^2 \xi_u^2
  e^{-\alpha\kappa\phi} \nonumber
  \\
  & \hspace{8.5mm}
   -\beta\kappa \left( V_a +\frac12 m^2 \xi^2 \right)
   e^{-\beta\kappa\phi}
  \bigg\} \bigg( \frac{\dot{a}}{a} \bigg)^{-2},
      \label{Eq-phi-S1}
  \\
  & \hspace{0mm} \rho_{\texttt{R}u} = - 4 \rho_\texttt{R}
  + \frac{\dot{a}}{a} \left( \Gamma_\xi \xi_u^2 + \Gamma_\phi \phi_u^2 \right),
      \label{Eq-rhoR-S1}
\end{align}
\end{subequations}
where the subscript $u$ denotes a derivative with respect to $u=-\ln(a_0/a)$ and $a_0\equiv a(u_0)$ represents the value of the scale factor at present $u_0=0$.

The quantities $\dot{a}/a$ and $\ddot{a}/a$, where an overdot denotes a derivative with respect to time $t$, are functions of $u$, $\xi$, $\xi_u$, $\phi$, $\phi_u$, and $\rho_\texttt{R}$ given by
\begin{align}
 \left( \frac{\dot{a}}{a} \right)^2 \equiv H^2=2\kappa^2
 \frac{ \left( V_a + \frac12 m^2 \xi^2 \right) e^{-\beta\kappa\phi} +
 \rho_\texttt{R} }{6 - \kappa^2 \phi_u^2
 - \kappa^2 \xi_u^2 e^{-\alpha\kappa\phi}},
 \label{Eq-dota-s1}
\end{align}
and, using $\ddot a/a = \dot H+H^2$,
\begin{align}
& \frac{\ddot{a}}{a} = \frac{\kappa^2}{3}
 \Bigg\{ \frac{2\kappa^2 \left[ \left( V_a+\frac12 m^2 \xi^2 \right)
 e^{-\beta\kappa\phi} + \rho_\texttt{R} \right](\phi_u^2 + \xi_u^2 e^{-\alpha\kappa\phi})}{\kappa^2 \phi_u^2 + \kappa^2 \xi_u^2 e^{-\alpha\kappa\phi} - 6}  \nonumber
 \\
& \hspace{6.4mm} + \left( V_a + \frac12 m^2 \xi^2 \right)
  e^{-\beta\kappa\phi} - \rho_\texttt{R}
    \Bigg\},
   \label{Eq-dotdota-s1}
\end{align}
where $H$ denotes the Hubble parameter.

The dissipation coefficients $\Gamma_\xi$ and $\Gamma_\phi$ are given by 
\begin{align}
 \Gamma_{\xi,\phi} = f_{\xi,\phi} \times \left\{
    \begin{aligned}
     & T^p,
     & T\geq T_\texttt{E},
    \\
     & T^p  \exp \left[ 1- \left( \frac{T_\texttt{E}}{T} \right)^q \right],
     & T\leq T_\texttt{E},
    \end{aligned}
 \right.
 \label{gammas}
\end{align}
where $T_\texttt{E}$ is the temperature of the radiation bath at the end of the inflationary period, $f_\xi$ and $f_\phi$ are positive constants with dimension $(\textrm{mass})^{1-p}$, and $q>0$ and $p$ are parameters, determining the temperature dependence of these coefficients.
Within this definition, we are not postulating any phenomenological or microscopic quantum field theory model, e.g.~\cite{Bastero-Gil:2016qru, Lima:2019yyv, Bastero-Gil:2019gao}.
Rather, it represents a model-independent approach with a generic temperature dependence, leading to an exponential suppression of the coefficients after the inflationary epoch.

To characterize the intensity of the energy transfer from the inflaton field to the radiation bath during the inflationary period, it is common to introduce the dissipation ratio $Q$, defined as $Q\equiv\Gamma/(3H)$. 
Depending on the value of $Q$, two regimes of warm inflation can be defined: the weak dissipative regime ($Q<1$) and the strong dissipative regime ($Q > 1$).
The two-scalar-field cosmological model under consideration can accommodate both regimes, as shown in Ref.~\cite{Sa:2020}. 
In this paper, we choose a set of values for the model's parameters and initial conditions (see below), yielding a strong dissipative regime.

At the end of the inflationary period, the dissipation coefficients $\Gamma_\xi$ and $\Gamma_\phi$ are exponentially
suppressed and, soon afterward, become negligible.
This marks the end of the first stage of evolution.
During the second stage, in the absence of dissipation, radiation decouples from the scalar field $\xi$, which begins to oscillate rapidly around the minimum of its quadratic potential\footnote{The oscillatory regime is ensured by the condition $M_\xi(\phi)\gg H$, where $M_\xi^2(\phi)\equiv m^2e^{-\beta\kappa\phi}$ is the effective mass of the inflaton field.}, behaving like a nonrelativistic dark-matter fluid with equation of state $p_\xi=0$ and energy density given by \cite{Sa:2020}
\begin{equation}
 \rho_\xi=Ce^{-3u}e^{\frac{(\alpha-\beta)}{a}\phi},
\end{equation}
where $C$ is a constant whose value is fixed by current cosmological measurements.
The previous expression reveals that in the presence of a direct coupling between the scalar fields $\xi$ and $\phi$, dark matter does not evolve like usual baryonic matter ($e^{-3u}$, or $a^{-3}$ in terms of the scale factor).
There is now an explicit dependence on the dark energy field $\phi$, requiring dark matter and baryonic matter to be treated separately.

Accordingly, during the second stage of evolution (radiation-, matter-, and dark energy-dominated eras), the evolution equations are

\begin{align}
 \phi_{uu} = {}& - \bigg\{
  \bigg[ \frac{\ddot{a}}{a} + 2 \bigg( \frac{\dot{a}}{a} \bigg)^2 \bigg] \phi_u
  -\beta\kappa V_a e^{-\beta\kappa\phi} \nonumber
\\
  & +\frac{(\alpha-\beta)\kappa C}{2}
  e^{\frac{(\alpha-\beta)\kappa}{2}\phi} e^{-3u} \bigg\} \left(
\frac{\dot{a}}{a} \right)^{-2}, \label{Eq-phi-s2}
\end{align}
with
\begin{align}
 \left( \frac{\dot{a}}{a} \right)^2= {}&
 2\kappa^2 \bigg[ V_a e^{-\beta\kappa\phi}
 + \left( \rho_{\texttt{BM}0}
 + C e^{\frac{(\alpha-\beta)\kappa}{2}\phi} \right) e^{-3u} \nonumber
\\
 & + \rho_{\texttt{R}0}e^{-4u} \bigg]
 \left( 6-\kappa^2 \phi_u^2 \right)^{-1},
\label{Eq-dota-s2}
\end{align}
and
\begin{align}
 \frac{\ddot{a}}{a} = {}& \frac{\kappa^2}{6} \bigg\{
 4\kappa^2 \bigg[ V_a e^{-\beta\kappa\phi}
     + \left( \rho_{\texttt{BM}0}
     + C e^{\frac{(\alpha-\beta)\kappa}{2}\phi} \right) e^{-3u} \nonumber
\\
 & + \rho_{\texttt{R}0} e^{-4u} \bigg]
 \phi_{u}^2 \left( \kappa^2 \phi_u^2 - 6 \right)^{-1}
 + 2 V_a e^{-\beta\kappa\phi} \nonumber
\\
 & - \left( \rho_{\texttt{BM}0}
   + C e^{\frac{(\alpha-\beta)\kappa}{2}\phi} \right) e^{-3u}
   - 2\rho_{\texttt{R}0} e^{-4u} \bigg\},
 \label{Eq-dotdota-s2}
\end{align}
where ordinary baryonic matter is described as a perfect fluid with pressure $p_\texttt{BM}=0$ and energy density $\rho_\texttt{BM} = \rho_{\texttt{BM}0} \,e^{-3u}$.

For the present values of the various Universe's components we consider some of the latest observational constraints \cite{Planck:2018vyg}, hence we adopt $\rho_{\texttt{R}0}=9.02\times10^{-128}\, m_\texttt{P}^4$, $\rho_{\texttt{BM}0}=8.19\times10^{-125}\, m_\texttt{P}^4$, $\rho_{\texttt{DE}0}=1.13\times10^{-123}\, m_\texttt{P}^4$, and $\rho_{\texttt{DM}0} = 4.25\times10^{-124}\, m_\texttt{P}^4$, yielding for the Hubble constant the value $H_0  = 1.17 \times 10^{-61} \, m_\texttt{P}$ or, in more familiar units, $H_0=67\,{\rm km}\,{\rm s}^{-1}{\rm Mpc}^{-1}$.

As a benchmark for our numerical constraints over the free parameters, we can follow Ref.~\cite{Luongo:2022}, where the two-scalar-field cosmological framework has been investigated by means of a Monte Carlo Markov chain (MCMC) analysis against low-redshift datasets, deriving constraints on the free parameters and comparing the model's performance to the standard $\Lambda$CDM scenario. 

Accordingly, it has been found that the condition $\lvert \alpha - \beta \rvert \lesssim 1$, turns out to be essential observationally and clearly introduces a limitation over the choice of the parameter space\footnote{The MCMC computation has been also explored adopting model selection criteria, such as the Akaike and Deviance information criteria, both indicating that our approach achieves a statistical fit comparable to the standard $\Lambda$CDM model.}. 

Hence, taking the bounds from the MCMC computation for granted, we employ
\begin{gather}
\alpha = 0.36^{+0.18}_{-0.26},
\label{v-alpha}
\\
\beta= 0.01^{+0.34}_{-0.24},
\label{v-beta}
\end{gather}
selecting for our base scenario the mean values, namely choosing $\alpha = 0.36$ and $\beta= 0.01$, with $H_0= 1.17 \times 10^{-61} \, m_\texttt{P}$, as well as $p=1$, $q=2$, $f_\xi=f_\phi=2$, $V_a=1.1\times10^{-123} \,m_\texttt{P}^4$ and $m = 10^{-5} \,m_\texttt{P}$.
In addition, the initial conditions have been selected to be $\xi_i=0.75 \, m_\texttt{P}$, $\phi_i=10^{-3} \,m_\texttt{P}$, $\xi_{u,i}=10^{-2} \,m_\texttt{P}$, $\phi_{u,i}=10^{-5} \,m_\texttt{P}$, and $\rho_{\texttt{R},i}=0.25\times10^{-12}\, m_\texttt{P}^4$. 
At the transition between the first and second stages of evolution, the different physical quantities are continuous.

\begin{figure}[t]
 \includegraphics[scale=1.3]{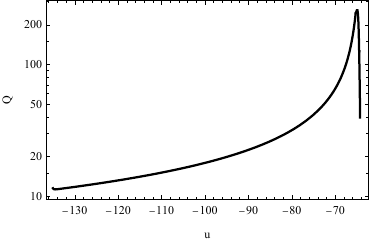}
\caption{Evolution of the dissipation ratio $Q$ during the inflationary period, which extends from $u\approx-135.3$ to $u\approx-65.3$ ($70$ e-folds of expansion), and the transition from inflation to the radiation-dominated era, which extends from $u\approx-65.3$ to $ u\approx-64.3$. Throughout the entire inflationary period, a strong dissipative regime is maintained. At the end of this period, the dissipation coefficients $\Gamma_\xi=\Gamma_\phi$ are exponentially suppressed, resulting in a sharp decrease of $Q$.}
\label{evolQ}
\end{figure}

\begin{figure}[t]
 \includegraphics[scale=1.3]{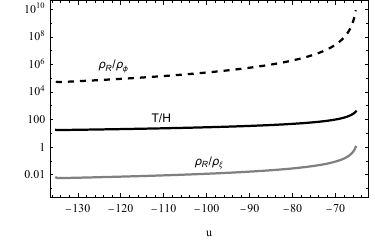}
\caption{Evolution of the ratios $T/H$, $\rho_\texttt{R}/\rho_\xi$, and $\rho_\texttt{R}/\rho_\phi$ during the inflationary period. The end of inflation occurs when $\rho_\texttt{R}/\rho_\xi=1$. From then on, radiation dominates the Universe's evolution.}
\label{evol-TH}
\end{figure}

The above choice of parameters and initial conditions corresponds to a strong dissipative regime ($Q>1$).
In Fig.~\ref{evolQ}, the evolution of $Q$ is shown for the entire inflationary period, as well as for the smooth transition from inflation to the radiation-dominated era, while Fig.~\ref{evol-TH} shows the evolution of the ratios $T/H$, $\rho_\texttt{R}/\rho_\xi$, and $\rho_\texttt{R}/\rho_\phi$ during inflation.

Finally, it is noteworthy to focus on the cosmological perturbations induced by such a model.
During the inflationary epoch, scalar and tensor perturbations are expected to be relevant and detectable.
The latter are studied in this article. As for the former, they will be analyzed in detail in a future publication; here, we will limit our comments to some general observations.
 Scalar perturbations are fundamental to validate cosmological models by comparing their predictions with the observational constraints provided by the Planck mission \cite{Planck:2018jri}. 
The two-scalar-field model under consideration presents a complex scenario in this regard; indeed, considering warm inflation within a multi-field approach leads to an unknown picture for describing scalar perturbations (as far as we know).
In warm inflation, the source of density fluctuations is the thermal fluctuations in the radiation bath, and this is a physically different scenario with respect to cold inflation, where the source is the quantum fluctuations of the inflaton field \cite{Lima:2019yyv, kamali:2023}.
On the other hand, multi-field approaches provide a more complex framework, where the interaction among the fields generally generates scalar perturbations that are superpositions of several contributions, such as entropic and adiabatic ones \cite{Langlois}.
Since both warm inflation and multi-field inflation have proven to be suitable frameworks, as shown in Refs.~\cite{Visinelli:2016rhn, Benetti:2016jhf, Montefalcone:2022jfw, B:2025koi, McAllister:2012am, Gong:2016qmq}, it is expected that the two-scalar-field cosmological model can be successfully constrained as well.
In particular, for the choice of parameters in the base scenario, since $\rho_\phi \ll \rho_\xi$ (see Fig.~\ref{evol-TH}), standard single-field warm inflation results should provide a good approximation.
Thus, as shown in Refs.~\cite{Benetti:2016jhf, Bastero-Gil:2016qru}, considering a strong dissipation regime should lead to observational consistency, with a reduced tensor-to-scalar ratio and a more blue-tilted spectrum as compared to cold inflation.
Nevertheless, to properly account for the predictions of the two-scalar-field cosmological model and to rigorously test its consistency, the generalization of the aforementioned approaches should be considered. 
However, this is not straightforward and, as mentioned above, is left for future work.


\section{Formalism of continuous Bogoliubov coefficients} \label{sezione3}

It is a consolidated fact that primordial gravitational waves can be generated along the Universe's expansion, resulting in a spectrum spanning a wide range of frequencies \cite{Grishchuk:1974, Starobinkii:1979, Abbott:1986, Allen:1988}.
These waves, being potentially detectable, open new avenues toward the cosmological early phases.

To calculate the corresponding gravitational wave spectrum, within the two-scalar-field cosmological model described in the previous section, we employ the formalism of \emph{continuous Bogoliubov coefficients}, quite well-established in the framework of field theories, in relation to different vacuum connections.

According to this formalism, the changes in the graviton creation and annihilation operators as the Universe evolves can be found by using Bogoliubov coefficients, defined as continuous functions of time\footnote{The continuous Bogoliubov coefficients differ from the well-known discrete Bogoliubov coefficients \cite{Herring:2019hbe}, since it is not necessary to assume any sudden transition between cosmological eras. Further, in cases where the two approaches can be both used, the results are found to agree.}.
In this case, the gravitons---the associated and so far hypothetical particles that constitute gravitational waves---are given in terms of these continuous coefficients, for any given epoch. 

The original formulation that makes use of continuous Bogoliubov coefficients was first applied in Ref.~\cite{Parker:69} with the aim of inferring particle production in an expanding Universe\footnote{Cosmological particle production is a widely studied phenomenon. In particular, this mechanism directly involves Bogoliubov transformations. For examples of particle production in different cosmological scenarios, see Refs. \cite{Ford:2021syk, Belfiglio:2025chv}. } and, afterwards, extended to the case of gravitons~\cite{Henriques:94, Moorhouse:94, Mendes:95}.
Several applications of this formalism have been developed within precise cosmological models, see e.g.~Refs.~\cite{Henriques:2007, Sa:2008, Sa:2009, Henriques:2009, Sa:2010, Sa:2012, Bouhmadi-Lopez:2010, Bouhmadi-Lopez:2013, Morais:2014}.

We now introduce gravitational waves in our prescription.
Thus, we take into account perturbations in the metric defined by Eq.~(\ref{metric}), specifically tensor perturbations.
The tensor perturbations $h_{ij}$ to the flat Friedmann-Lema\^{i}tre-Robertson-Walker metric
\begin{equation}
    ds^2=a^2(\eta) \{ -d\eta^2 + [ \delta_{ij} + h_{ij}(\eta,\textbf{x}) ] dx^i dx^j \},
\end{equation}
where $\eta$ is the conformal time\footnote{In this section, conformal time is adopted as the temporal coordinate in order to simplify the presentation.
At the end, the results are reformulated in terms of the variable $u$, used in the evolution equations of the two-scalar-field cosmological model.}, can be expanded in terms of plane waves, 
\begin{equation}
\begin{split} h_{ij}(\eta,\textbf{x})=\kappa\sum_{p=1}^2 &\int \frac{d^3k}{(2\pi)^{3/2}a(\eta) \sqrt{2k}}\left[a_p(\eta,\textbf{k}) \right. \\
     &\quad\,\left.\times \varepsilon_{ij}(\textbf{k},p) e^{i\textbf{k}\cdot\textbf{x}}\chi(\eta,\textbf{k}) + \mbox{h. c.}\right]
\end{split}
\end{equation}

In the previous expression, $\textbf{x}$ is the spatial-coordinates three-vector, $\textbf{k}$ is the comoving wave-number three-vector, such that $k=|\textbf{k}|=a\, \omega$, $\omega$ is the angular frequency, $p$ runs over the two polarizations of the gravitational waves, $a_p$ is the annihilation operator, $\varepsilon_{ij}$ is the polarization tensor, and the mode function $\chi$ satisfies the equation of a parametric oscillator
\begin{equation}
    \chi^{\prime\prime} +\left( k^2-\frac{a^{\prime\prime}}{a} \right) \chi = 0,
\end{equation}
where a prime denotes a derivative with respect to the conformal time $\eta$.

In an expanding Universe, the annihilation $a_p(\eta,\textbf{k})$ and creation $a_p^\dagger(\eta,\textbf{k})$ operators change over time, and such change is codified in terms of the time-fixed annihilation $A_p(\textbf{k})$ and creation $A_p^\dagger(\textbf{k})$ operators through the Bogoliubov transformation
\begin{equation}
a_p(\eta,\textbf{k})=\alpha_k(\eta,k)A_p(\textbf{k})+\beta_k^*(\eta,k)A_p^\dagger(\textbf{k}),
\end{equation}
where the Bogoliubov coefficients $\alpha_k$ and $\beta_k$ satisfy the condition $|\alpha_k|^2-|\beta_k|^2=1$ and the number of gravitons created is given by $\langle N_k(\eta) \rangle = |\beta_k|^2$.
For a well-defined framework and a proper introduction of the Bogoliubov transformations, the initial and final vacuum states for the graviton must be specified.
We take the Bunch–Davies vacuum \cite{Bunch:1978yq} as the initial state, while at the end of its evolution the graviton satisfies the adiabatic conditions \cite{Henriques:94}.

To compute the Bogoliubov coefficients, one solves the system of ordinary differential equations
\begin{subequations}\label{Bogoliubov eqs}
\begin{align}
    \alpha_k^\prime &= \frac{i}{2k} \left[ \alpha_k+\beta_k e^{2ik(\eta-\eta_i)}\right] \frac{a^{\prime\prime}}{a}, \\
    \beta_k^\prime  &= -\frac{i}{2k} \left[ \beta_k+\alpha_k e^{-2ik(\eta-\eta_i)}\right] \frac{a^{\prime\prime}}{a},
\end{align}
\end{subequations}
with initial conditions $\alpha_k=1$ and $\beta_k=0$, corresponding to the absence of gravitons at the beginning of the Universe's evolution.
Under appropriate circumstances, this system of equations may admit exponentially growing solutions corresponding to graviton production.

Upon the redefinition
$\alpha_k=\frac12(X+Y)\exp\{ik(\eta-\eta_i)\}$ and $\beta_k=\frac12(X-Y)\exp\{-ik(\eta-\eta_i)\}$, the above system of equations becomes
\begin{subequations} \label{SODE-XY}
\begin{align}
    X^\prime&=-ikY, \label{Xprime}
\\
    Y^\prime&=-\frac{i}{k} \left( k^2-\frac{a^{\prime\prime}}{a} \right) X, \label{Yprime}
\end{align}
\end{subequations}
with initial conditions $X_i=Y_i=1$.
The number of gravitons created is now given by $\langle N_k(\eta) \rangle = |\beta_k|^2 = (X-Y)(X^*-Y^*)/4$.
Of course, to solve this system of equations, it is necessary to specify a model of evolution of the Universe for the entire period under consideration in order to determine $a^{\prime\prime}/a$.

For $k^2\lesssim a^{\prime\prime}/a$, the system of ordinary differential equations (\ref{SODE-XY}) describes the production of gravitational waves with angular frequency $\omega=k/a$, while for $k^2 \gg a^{\prime\prime}/a$, it describes a harmonic oscillator, implying that no gravitational waves are produced.
This circumstance will be utilized in our numerical calculations (see Sec.~\ref{GW spectrum}) to reduce the computation time of gravitational wave spectra substantially.

The gravitational-wave spectral energy density parameter $\Omega_{\texttt{GW}}$ is defined as
\begin{equation}\label{eq spectrum}
\Omega_{\texttt{GW}}(\omega_0)=\frac{8\hbar G}{3\pi c^5 H^2_0} \omega_0^4 \,(|\beta_{k}|^2)_0,
\end{equation}
where $\hbar$ is the reduced Planck constant, $c$ is the speed of light, and the subscript zero denotes quantities evaluated at the present time\footnote{We use the natural system of units ($\hbar=c=1$) to study the evolution of the Universe within the two-scalar-field cosmological model described in the previous section.
However, following a common practice, we present $\Omega_{\texttt{GW}}$ in units of the International System.}.

To apply the above described formalism of continuous Bogoliubov coefficients to the two-scalar-field cosmological model under consideration, we rewrite the system of equations~(\ref{SODE-XY}) in terms of the variable $u$ introduced in the previous section, namely, 
\begin{subequations} \label{SODE-XYu}
\begin{align}
    &X_u=-i\omega_0 e^{-u} \frac{Y}{\dot{a}/a}, \label{SODE-XYua}
\\
    &Y_u=-\frac{i}{\omega_0} e^u \left[ \omega_0^2 e^{-2u} -\frac{\ddot{a}}{a}-\left( \frac{\dot{a}}{a} \right)^2 \right] \frac{X}{\dot{a}/a}, \label{SODE-XYub}
\end{align}
\end{subequations}
where $\ddot{a}/a$ and $\dot{a}/a$ are given by Eqs.~(\ref{Eq-dota-s1}) and (\ref{Eq-dotdota-s1}) for the first stage of evolution, and Eqs.~(\ref{Eq-dota-s2}) and (\ref{Eq-dotdota-s2}) for the second stage.


\section{Gravitational-wave spectrum} \label{GW spectrum}

Using the formalism of continuous Bogoliubov coefficients, we now calculate the full energy spectrum of the gravitational waves generated within the two-scalar-field cosmological model described in Sec.~\ref{2SF model}.

To this end, we numerically solve the system of equations~(\ref{SODE-XYu}) for a specific value of $\omega_0$, with initial conditions corresponding to the absence of gravitational waves ($X_i = Y_i = 1$).
In this system of equations, the values of $\dot{a}/a$ and $\ddot{a}/a$ as functions of $u$ are known from the beginning of inflation to the present moment (see Sec.~\ref{2SF model}).
In this way, we obtain $X$ and $Y$ as functions of $u$, which, in turn, allows us to obtain $|\beta_k|^2$ as a function of $u$, and, in particular, its value at the present moment, $(|\beta_k|^2)_0$.
Finally, using Eq.~\eqref{eq spectrum}, we calculate $\Omega_{\texttt{GW}}$ for the value of $\omega_0$ under consideration.
Repeating this procedure for multiple values of $\omega_0$ between $\omega_\texttt{min}$ and $\omega_\texttt{max}$, we obtain the full energy spectrum of gravitational waves.

\begin{figure}[t]
 \includegraphics[scale=1.3]{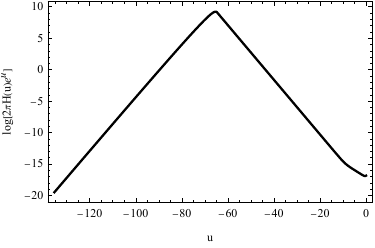}
\caption{The minimum angular frequency of a gravitational wave corresponds to a wavelength equal, today, to the Hubble distance, $\omega_\texttt{min} = 2\pi c/d_\texttt{Hub}(u_0)e^{u_0} \approx 2\pi H_0=10^{-16.86}\, \mbox{rad/s}$, where the present value of the Hubble parameter $H_0\equiv H(u_0=0)$ is taken to be $2.17\times10^{-18}\, \mbox{s}^{-1}$ in our base scenario. The maximum angular frequency corresponds to a wavelength equal to the Hubble distance at the end of the inflationary period $u_\texttt{inf}$, afterwards redshifted by the expansion of the Universe, $\omega_\texttt{max} \approx 2\pi H(u_\texttt{inf}) e^{u_\texttt{inf}} = 10^{9.20} \, \mbox{rad/s}$ for the base scenario.}
\label{min-max-freq}
\end{figure}

The angular frequency takes values ranging from about $10^{-17}$ rad/s to about $10^9$ rad/s.
These frequencies correspond to wavelengths equal to the Hubble distance today and at the end of the inflationary period, respectively (see Fig.~\ref{min-max-freq}). 

The procedure described above to obtain the energy spectrum of gravitational waves can be greatly simplified by considering that, in Eq.~(\ref{SODE-XYub}), when the term  $\omega_0^2e^{-2u}$ is much larger than the term $|\ddot{a}/a+(\dot{a}/a)^2|$, gravitational waves are not produced [see discussion after Eq.~\eqref{Bogoliubov eqs}].
In other words, the equations describing the Bogoliubov coefficients lead to two regimes: a harmonic oscillating regime without graviton production, and another with graviton production.
Thus, in practice, before starting the computation of $X$ and $Y$, we first compare the above terms for each value of $\omega_0$, identify the intervals of $u$ for which gravitational waves are not produced, and exclude these intervals from the numerical computation.

Adopting this simplified procedure, in the base scenario, in the angular frequency range between $\omega_\texttt{min} = 10^{-16.86} \, \mbox{rad/s}$ and $\omega_0\approx10^{-15}\, \mbox{rad/s}$, the numerical calculation is performed for the entire range of $u$, from the beginning of the inflationary period to the present time.
However, for frequencies above $10^{-15} \, \mbox{rad/s}$ and up to the maximum frequency $\omega_\texttt{max} = 10^{9.20} \, \mbox{rad/s}$, numerical computations begin later, shortly before the harmonic oscillation regime ends, and, at the final stages of evolution, they end earlier, shortly after this regime starts again.
That is, the higher the frequency value considered, the higher the value of $u$ for which computations begin and the lower the value of $u$ for which they end.
To further reduce computation time, for frequencies greater than approximately $10^{-10} \, \mbox{rad/s}$, noting that the generation of gravitational waves in all late eras of the Universe's evolution is utterly negligible, numerical computations are stopped shortly after the beginning of the radiation-dominant era.

\begin{figure}[t]
 \includegraphics[scale=1.3]{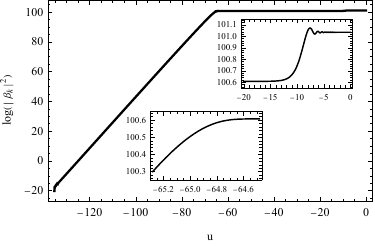}
\caption{Evolution of $|\beta_k|^2$ as a function of $u$ for the base scenario and $\omega_0=10^{-16}$ rad/s. During inflation, which lasts from $u=-135.3$ to $u=-65.3$, gravitational waves are generated copiously. During the transition from inflation to the radiation-dominated era (lower inset) and, as well as, during the transition to a matter-dominated and then dark energy-dominated eras (upper inset), more gravitational waves are produced, but incomparably fewer than during the inflationary period. During the radiation-dominated era, no gravitational waves are produced.}
\label{beta2}
\end{figure}

Our numerical computations reveal that most gravitational waves are produced during the primordial inflationary period.
Additional gravitational waves are generated during the transition from inflation to the radiation-dominated era and, for low frequencies, also during the transition from the latter to the matter- and dark energy-dominated eras.
However, the amount of gravitational waves produced during these transitions is incomparably less than those generated during inflation (see Fig.~\ref{beta2}).

\begin{figure*}[t]
 \includegraphics[scale=1.3]{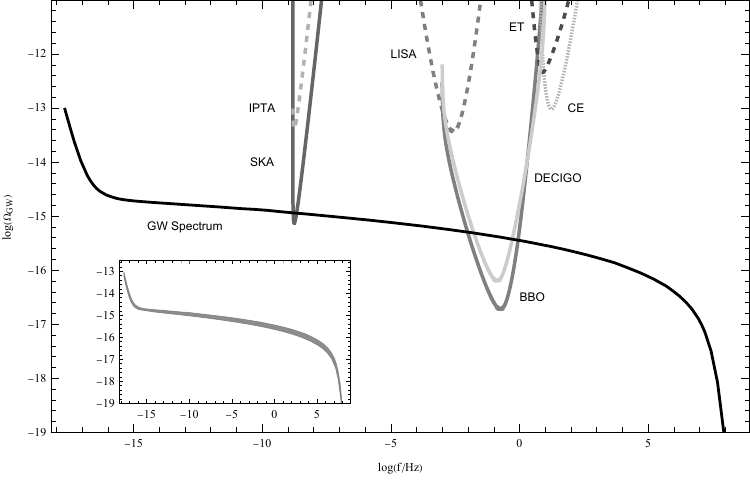}
\caption{Full gravitational-wave energy spectrum for the base scenario of the two-scalar-field cosmological model, superimposed on the sensitivity curves of planned next-generation ground- and space-based gravitational-wave detectors, LISA, BBO, CE, ET, SKA, IPTA, and DECIGO \cite{Schmitz:2020syl} (for the repository with the sensitivity curves see Ref.~\cite{repository}). While the main plot displays the energy spectrum obtained for the values of $\alpha$ and $\beta$ adopted in the base scenario, the inset shows the envelope containing the spectra generated when the error bars for these parameters are taken into account (see Eqs.~\eqref{v-alpha} and \eqref{v-beta}).}
\label{spectrum-base}
\end{figure*}

In Fig.~\ref{spectrum-base}, we present the full gravitational-wave energy spectrum for the base scenario of the two-scalar-field cosmological model, whose parameter values and initial conditions were specified at the end of Sec.~\ref{2SF model}. 
The gravitational wave density parameter $\Omega_{\texttt{GW}}$ is presented as a function of the frequency $f$, related to the angular frequency $\omega_0$ by the relation $\omega_0=2\pi f$.
The sensitivity curves of planned ground- and space-based gravitational-wave detectors---Laser Interferometer Space Antenna (LISA), Big Bang Observer (BBO), Cosmic Explorer (CE), Einstein Telescope (ET), Square Kilometre Array (SKA), International Pulsar Timing Array (IPTA), and Deci-Hertz Interferometer Gravitational Wave Observatory (DECIGO)---are superimposed on the spectrum\footnote{For further details regarding the sensitivity curves and characterization of the detectors, see Refs.~\cite{Moore:2014lga, Ringwald:2020vei}}.

\begin{figure*}[t] 
   \includegraphics[scale=1.3]{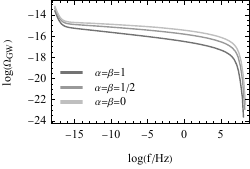}
  \includegraphics[scale=1.3]{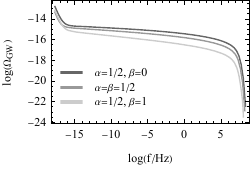}
  \includegraphics[scale=1.3]{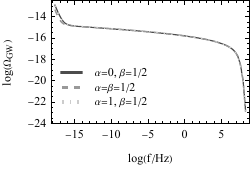}
   \includegraphics[scale=1.3]{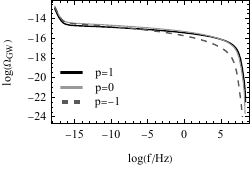}
 \includegraphics[scale=1.3]{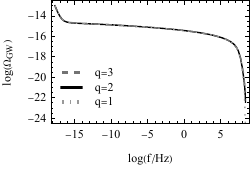}
  \includegraphics[scale=1.3]{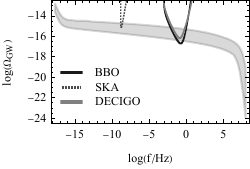}
\caption{Full gravitational-wave energy spectra for different parameter choices. In the upper row, the plots show the spectra for the two-scalar-field cosmological model with free parameters $\alpha$ and $\beta$: (i) equal, (ii) $\alpha$ fixed, and (iii) $\beta$ fixed. In the lower row, we present the spectra obtained for different choices of the temperature dependence in the dissipation coefficients, where the parameters $p$ and $q$ are varied. Finally, the envelope of all the spectra corresponding to the considered set of parameters is shown, superimposed on the sensitivity curves of BBO, SKA, and DECIGO. }
\label{y}
\end{figure*}

From this figure, we can readily conclude that, within the two-scalar-field cosmological model, there is a promising frequency range for the detection of primordial gravitational waves by future detectors, specifically between $10^{-2}$ and $1$ Hz, which will be accessible to BBO and DECIGO.
At lower frequencies, around $10^{-9}$ to $10^{-8}$ Hz, the prospects are less favorable, as the SKA sensitivity curve only marginally intersects the energy spectrum.

These conclusions are reinforced by an analysis of the gravitational-wave energy spectra obtained for various values of the relevant model parameters, $\alpha$, $\beta$, $p$, and $q$ (see Fig.~\ref{y}).
Across the entire frequency range, these spectra are qualitatively similar to the spectrum obtained in the base scenario and, quantitatively, differ by less than about an order of magnitude in $\Omega_{GW}$. 
Specifically, as the spectrum is mainly due to the inflationary expansion, the parameters that significantly affect the magnitude of $\Omega_{GW}$ are $\beta$ and $p$.
Indeed, they directly influence inflation, since the former characterizes the dressed mass of the inflaton, while the latter determines the rate of energy transfer from the inflaton to the radiation bath.
On the other hand, $\alpha$ is a kinetic coupling strength and $q$ governs the transition from inflation to the radiation-dominated era, and as a consequence, they have virtually no effect on the production of gravitational waves.
Figure~\ref{y} (bottom right plot) also displays the envelope containing all the spectra corresponding to the different sets of parameters considered, superimposed on the sensitivity curves of the planned gravitational wave detectors BBO, DECIGO, and SKA.

Notably, it should be emphasized that, although we adopted for the present value of the Hubble parameter $H_0 = 67\,{\rm km}\,{\rm s}^{-1}{\rm Mpc}^{-1}$, our physical results are general.
Considering different surveys and, as a consequence, assuming for example $H_0 = 73\,{\rm km}\,{\rm s}^{-1}{\rm Mpc}^{-1}$, would lead to a final gravitational-wave spectrum differing from the initial one by a few percentage points, but not significantly enough to affect the physical interpretation.

As a general remark, the detectability of gravitational waves in two-scalar-field cosmological models, in contrast to single-field inflation, where tensor modes are dominantly vacuum fluctuations, introduces additional dynamical degrees of freedom, enabling new gravitational wave production channels such as parametric resonance during reheating \cite{Kofman:1997yn}, isocurvature-to-curvature conversion \cite{Lyth:2001nq}, and nonlinear emission derived from quasi-particles, such as oscillons and solitons \cite{Copeland:1995fq}, geometric particles \cite{Belfiglio:2022qai, Belfiglio:2023rxb, Belfiglio:2024swy, Belfiglio:2024xqt} and so on. 

Hence, although predictions are computationally demanding and constrained by other cosmological data, the diversity of spectral features and possible multi-band detections has been simplified here in order to make reasonable predictions on the most viable bounds expected by the presence of such a non-minimal coupled Lagrangian. 

Future attempts will clearly shed light on refinements and more accurate constraints to clarify the role of the fields under examination, even tuning different bands for the free parameters involved in our computation.


\section{Final outlooks}\label{conclusions}

In this work, we studied the gravitational-wave background predicted within a two-scalar-field cosmological framework, initially proposed to unify the mechanism of inflation, with the existence of dark matter and dark energy,  under the same standards. 

The underlying model we studied originated from an action inspired by a variety of extended theories of gravity, including, among all, hybrid metric-Palatini gravity, Jordan-Brans-Dicke theory, Kaluza-Klein compactifications, $f(R)$ gravity, and low-energy string-inspired formulations. 

The formal scheme consists of non-minimally coupling two scalar fields and recovering, at different energy scales, the effects either due to inflation or to the dark sector.
The paradigm is thus characterized by a non-minimal scalar sector, while both fields are minimally coupled to the Ricci scalar, providing a flexible yet robust structure to capture both early- and late-time cosmic dynamics, without altering the gravitational constant. 

We postulated a motivated effective potential for one of the scalar fields, ensuring a contribution provided by a bare mass term, and derived the full set of cosmological equations of motion, from which we tracked the background evolution across all cosmological epochs.
Based on these equations, we computed the gravitational-wave spectrum spanning wavelengths from the present Hubble radius down to scales associated with the end of inflation. 

To do so, we performed a computation, using the formalism of continuous Bogoliubov coefficients, in analogy to the works of gravitational particle productions, where quantum particle production in the expanding Universe was presented and, here, in particular, extended to the case of graviton production. 

Indeed, through this formalism, we were able to follow the creation and annihilation of gravitons throughout cosmic history, thus capturing the continuous evolution of the gravitational-wave background across different cosmic eras.  

Our results demonstrate that the inflationary epoch was responsible for producing the dominant fraction of the gravitational-wave background, with additional but significantly weaker contributions generated during the transitions between inflation and radiation domination, radiation and matter domination, and finally the onset of the dark-energy-dominated epoch. 

Nevertheless, these subdominant contributions, while relatively suppressed, offer potential observational signatures of post-inflationary processes.  

Hence, we presented the energy density spectrum of gravitational waves, $\Omega_{\mathrm{GW}}(f)$, as a function of frequency and systematically compared it with the projected sensitivities of future ground- and space-based detectors, including LISA, BBO, CE, ET, SKA, IPTA, and DECIGO. 

This comparison highlighted promising frequency regions and bounds in which the predictions of the two-scalar-field cosmological model could, in principle, be tested.
Such comparisons emphasized that primordial gravitational-wave signals could serve as a powerful observational channel to test extensions of general relativity and multi-field inflationary scenarios.  

The results strengthened the case for multi-field approaches to gravitational waves whose signature, derived here,  highlighted the potential role of next-generation gravitational-wave detectors in discriminating between competing models of the early Universe and constraining the mechanisms of inflation and those behind the dark sector existence. 

Future works may extend the present investigation, noticing the role played by our scalar field in predicting small perturbations and, accordingly, the large-scale structures.
Moreover, we intend to explore even broader classes of potentials and couplings between the scalar fields, incorporating reheating dynamics in more detail, and performing parameter-space analyses via cosmological data.
Further improvements could also involve the inclusion of higher-order quantum corrections and non-Gaussian signatures in the gravitational-wave spectrum, thus shedding light on the connection between gravitational-wave astronomy and multi-field cosmology.


\begin{acknowledgments}
OL acknowledges support by the  Fondazione  ICSC, Spoke 3 Astrophysics and Cosmos Observations. National Recovery and Resilience Plan (Piano Nazionale di Ripresa e Resilienza, PNRR) Project ID $CN00000013$ ``Italian Research Center on  High-Performance Computing, Big Data and Quantum Computing" funded by MUR Missione 4 Componente 2 Investimento 1.4: Potenziamento strutture di ricerca e creazione di ``campioni nazionali di R\&S (M4C2-19)" - Next Generation EU (NGEU).
TM acknowledges hospitality to the University of Algarve during the time in which this paper has been finalized. TM is also grateful to Youri Carloni for fruitful discussions.
PS acknowledges support from CIDMA-Center for Research and Development in Mathematics and Applications under the Portuguese Foundation for Science and Technology Multi-Annual Financing Program for R\&D Units, grants UID/04106/2025 (\url{https://doi.org/10.54499/UID/04106/2025}) and UID/PRR/04106/2025 (\url{https://doi.org/10.54499/UID/PRR/04106/2025})
\end{acknowledgments}


\end{document}